# Determination of the Local Symmetry and the Multiferroic-ferromagnetic Crossover in $Ni_{3-x}Co_xV_2O_8$ by using Raman Scattering Spectroscopy

Yu-Seong Seo, Sun-Hwa Kim, and Jai Seok Ahn[*]

Department of Physics, Pusan National University, Busan 609-735, Republic of Korea

Il-Kyoung Jeong

Department of Physics Education, Pusan National University, Busan 609-735, Republic of Korea

Comprehensive vibrational studies on the multiferroic-to-normal-ferromagnetic crossover in isostructural $Ni_{3-x}Co_xV_2O_8$ (NCVO, $x = 0 - 3.0$) are performed using Raman scattering spectroscopy. The systematically red-shifted phonon modes are discussed in terms of the mode Grüneisen parameters, and are interpreted as a chemical pressure effect. In addition, we present evidence that the local symmetry is broken in the multiferroic phase.

## I. INTRODUCTION

Magnetism and ferroelectricity may coexist and interact with each other in multiferroic materials. Research on the coupling between magnetic and dielectric properties has become important in recent years for new fundamental phenomena and for new source technologies [1, 2]. Multiferroic materials are often found in geometrically frustrated magnetic materials, for example $RMnO_3$ and $RMn_2O_5$ ($R$ = rare earth atom), $CuFeO_2$, $CoCr_2O_4$, $Ni_3V_2O_8$, and hexagonal ferrites [3-9].

$M_3V_2O_8$ ($M$ = Ni or Co) is a Kagomé-staircase compound [10]. In this case, the ideal triangular spin-Kagomé planes buckle to form a three-dimensional staircase-like structure. The complex magnetic structures of $Ni_3V_2O_8$ (NVO) were analyzed with a model (spin-1 antiferromagnet on a Kagomé-staircase) including competing spin interactions, i.e., the nearest-($J_1$)-, the next-nearest-($J_2$)-interaction, and smaller anisotropy terms including the Dzyaloshinskii-Moriya (DM) interaction [11, 12]. In contrast, for $Co_3V_2O_8$ (CVO), spins interact ferromagnetically along the spine-spins, and the $J_3$ interaction between the spines governs its ferromagnetic features [13, 14]. Concerning the mixed system $Ni_{3-x}Co_xV_2O_8$ (NCVO), the magnetic properties were investigated by Qureshi *et al*. [15, 16]. In NVO, ferroelectricity develops with an incommensurate magnetic phase: a spontaneous polarization (~ 100 $\mu C/m^2$) lies along the *b*-axis and it is found to be proportional to the incommensurate magnetic order parameters in the Landau model [8]. According to neutron scattering and infrared measurements, the infrared-active $B_{2u}$ phonon modes ($b_4$ and $b_9$) induce a significantly large dipole moment along the *b*-axis and these modes are assumed to be responsible for the observed spontaneous polarization [17-19]. For Raman-active phonon modes, only limited information is available from an earlier work by Sudakar *et al*. on NVO [20].

In this work, we analyzed phonon modes from Raman scatterings of NCVO samples. The Raman spectrum of NVO was closely examined with previous neutron data [17] and infrared data [18, 19]. Phonon modes show a systematic softening behavior as a function of Co substitution, which is interpreted as a chemical pressure effect. Certain modes disappear as Co substitutes at Ni sites, demonstrating a local symmetry change around $x \simeq 2.0$. From this local symmetry difference, we obtain a picture of a multiferroic-ferromagnetic crossover in the NCVO system.

## II. EXPERIMENTAL

High-quality polycrystalline $Ni_{3-x}Co_xV_2O_8$ ($x$ = 0, 0.5, 1.0, 1.5, 2.0, 2.5, and 3.0) specimens were used for the present study. The samples were prepared by using a solid-state reaction method with stoichiometric amounts of NiO (Aldrich, 99.99%), $Co_3O_4$ (Aldrich, 99.99%), and $V_2O_5$ (Alfa Aesar, 99.99%). The samples were sintered at 1023, 1073, and 1123 K for more than 48 hours in air, with intermediate grindings. High-cobalt-concentration samples ($x \geq 1.5$) were prepared using argon atmosphere to optimize the oxygen content.

To check the phase-formation, we performed x-ray diffraction (XRD) $\theta$-$2\theta$ measurements at room temperature by using a Cu $K_\alpha$ line with a wavelength of 1.5406 Å (Bruker, D8 Advance with a Ni filter and a Lynxeye detector). No secondary phases were detected in the diffraction patterns. The results of the XRD analysis indicated that all the samples were in a single phase with an orthorhombic structure. As to space group, the structure belongs to *Cmca*, in accordance with Qureshi *et al*.'s report [15]. The XRD patterns show systematic peak shifts as a function of Co concentration. In addition, Ni-rich samples show more pronounced peak splittings than Co-rich ones. The crystal structures of NCVO series were refined from the XRD patterns by using the Rietveld method provided by RIETAN-2000 [21]. The structures were found to be orthorhombic (space group *Cmca*) with lattice constants of $a = 5.9268$, $b = 11.3766$, and $c = 8.2332$ Å for NVO, and $a = 6.0344$ $b = 11.4897$, and $c = 8.3001$ Å for CVO. The lattice parameters are similar to those in previous reports [10, 15, 22]. The effective ionic radius of $Ni^{2+}$ ($3d^8$ with spin-1) is 0.690 Å in the octahedral coordination. As for $Co^{2+}$ ion ($3d^7$), it is 0.65 and 0.745 Å for low-spin and high-spin configurations, respectively [23]. Because the cell size of CVO is the larger than that of the NVO, the $Co^{2+}$ ion in NCVO has a high-spin configuration (with spin-3/2). Note that the unit cell volume, $V$, increases almost linearly with $x$ in conformity with the Vegard's law [24]. Therefore, the mixed system NCVO can be thought of as forming a solid solution.

The infrared (IR) transmission spectra of NVO and CVO were obtained using a standard KBr pellet method in the mid-IR range. The IR spectra were recorded using a Fourier transform infrared spectrophotometer (Jasco-6100 with KBr beam splitter and DLATGS detector). Samples were pressed into pellets with concentrations of 0.1, 0.5, and 1.0 wt.%, diluted with KBr powders. The IR spectra were collected in the energy range of 350-4000 $cm^{-1}$. The IR transmission spectra of NVO and CVO are presented in Fig. 1. The IR spectra of NVO show absorption modes at 425 ($B_{3u}$), 458 ($B_{2u}$), and 659 ($B_{2u}$), and triplets of 795 ($B_{1u}$ and $B_{3u}$), 829 ($B_{1u}$), and 885 $cm^{-1}$ ($B_{2u}$), in agreement with the literature values [18]. CVO shows similar absorption structures at 648 ($B_{2u}$) and triplets of 774 ($B_{1u}$ and $B_{3u}$), 821 ($B_{1u}$), and 877 $cm^{-1}$ ($B_{2u}$). The four modes of CVO are red-shifted from the corresponding modes of NVO.

The red-shift of the four modes is explained with the volume expansion effect, which is associated with the mode Grüneisen parameter $\gamma_{\vec{k}s} = -\partial \ln \omega_s(\vec{k}) / \partial \ln V$ [25]. Here, $\omega_s(\vec{k})$ is the frequency of the phonon mode with a wavevector $\vec{k}$ and $V$ is the volume. We obtained the mode Grüneisen parameters for the four red-shifted modes, 0.456, 0.721, 0.264, and 0.247. These values are small compared to the value of ~ 1.5 for ionic crystals at room temperature [25]. Note, however, that here we are using two different materials in calculating the Grüneisen parameter, such that it is not from a thermal expansion in one system; rather, it is from a chemical pressure effect caused by ionic substitution.

The Raman spectra were obtained with standard confocal optics: a confocal sampling chamber (Dongwoo Optron, Vector-01FX True Confocal Raman), a monochromator (Dongwoo Optron, $f$ = 50 cm) equipped with a notch filter (OptoSigma), and a CCD camera (Andor, DV401A). An ×50 objective (Sigma Koki, S Plan Apo HL) was used both to focus and to collect the light in a back-scattering geometry. As an excitation source, the 488.0 nm line of an argon-ion laser (Coherent, Innova-70c) was used. The laser power at the focal spot (2-3 $\mu$m in diameter) was kept below 20 mW to prevent overheating. The light power was varied from 8 to 18 mW by using a graded attenuation filter. Raman spectra of NCVO were obtained at a laser power of 18 mW, except for NVO (8 mW). The low- and the high-frequency regions were measured with different gratings.

## III. RESULTS AND DISCUSSION

The Raman scattering result for NVO in Fig. 2(a) shows 12 weak modes, with the single strongest one being at ~ 826 cm$^{-1}$. We indexed the dominant modes with numbers (#1 to #8) and the small satellite modes with alphabets (*a* to *e*). According to the group theoretical analysis [18], there are 78 Γ-point phonon modes. Among the 78 modes, the number of Raman active modes is 36 ($10A_g + 8B_{1g} + 7B_{2g} + 11B_{3g}$).

In the polyhedral construction, NVO is composed of one type of VO$_4$ tetrahedron, with V-O bond lengths of 1.700, 1.700, 1.717, and 1.808 Å, and two types of NiO$_6$ octahedra. The single strongest mode, #8, resembles the Raman spectrum of an ideal VO$_4$ molecular block, as observed in BiVO$_4$ [26]. Therefore, peak #8 (~ 826 cm$^{-1}$) can be assigned as the symmetric V-O stretching mode with Raman active $A_g$ symmetry, as provided by Hardcastle *et al*. [26]. Note that the NiO$_6$ block is heavier than the VO$_4$ molecule; thus, the frequencies of the Raman modes from NiO$_6$ should be small compared with peak #8.

The Raman spectrum in Fig. 2(a) is very comparable to the inelastic neutron scattering (INS) spectrum of Harris *et al*. [17] in Fig. 2(b). The INS spectrum is more intense in the low-energy region below 500 cm$^{-1}$ than the Raman spectrum. For comparison, the positions of IR active phonons are depicted with lines in Fig. 2(b) [18]. Interestingly, the INS modes at ~ 649 cm$^{-1}$ (indexed with *β*) and at ~ 800 cm$^{-1}$ are similar to our two strongest Raman modes, #6 and #8, respectively. If we look more closely, we can see that the doublet modes, #6 (at 642 cm$^{-1}$) and #7 (at 676 cm$^{-1}$), correspond to the former INS *β*-peak. The weight-averaged frequency of the doublet modes are calculated as ~ 656 cm$^{-1}$. In addition, they are close to the IR active $B_{2u}$ mode ($b_9$) at ~ 659 cm$^{-1}$.

General IR and Raman wisdom is that they measure different symmetries: IR

measures polar modes (*ungerade*-) and Raman observes symmetric vibrations (*gerade*-). Also, note that INS is insensitive to such symmetry. Therefore, it is very intriguing that Raman spectroscopy observed doublet modes near the IR $B_{2u}$ mode ($b_9$), previously interpreted as being one of the important modes for spontaneous polarization in multiferroic NVO [17-19]. In this respect, our observation of the doublet modes may provide new insight into the theoretical understanding of multiferroicity of NVO.

The Raman spectra of NCVO change systematically with increasing Co-concentration, *x*, as presented in Fig. 3. The primary Raman modes are observed below ~ 1000 cm$^{-1}$ in Fig. 3(a), and their frequencies are tabulated in Table 1. Most of the modes shift to lower energy as Co substitutes at the Ni site. The frequencies of the dominant modes, $\omega_j$'s (of #2, #3, #6, and #8), are displayed in Fig. 3(c). They show a monotonic red-shift as a function of *x*, and the red-shift is linearly related to the volume expansion ($\Delta V$), as shown in Fig. 3(d). From this relation, the mode Grüneisen parameters of each mode can be directly evaluated by using $\gamma_j = -(\Delta\omega_j/\omega_j)/(\Delta V/V)$. The mode Grüneisen parameters ($\gamma_j$) are also presented in Table 1.

Certain modes disappear as Co substitutes at the Ni site. Small satellite peaks are observed at ~ 182, 256, and 323 cm$^{-1}$ (indexed as *a*, *b*, and *d*) in *x* = 0 (NVO); however, they rapidly merge to their nearest Raman modes as *x* increases. In addition, peak *c* (~ 270 cm$^{-1}$) almost disappears for *x* > 2.0. We note that mode *c* is close to the INS mode *α* in Fig. 2(b) and it is close to the IR $B_{2u}$ mode ($b_4$), which is interpreted as one of the important modes for the ferroelectricity of NVO [17-19]. From these, we can infer that the missing multiferroicity in CVO can be explained by the missing modes. Therefore, the local symmetry of NVO seems to be different from that of CVO, even though their global symmetries are the same (*Cmca*). We found similar claims of local symmetry change in (Mn,Co)WO$_4$ [27].

The high-frequency part of the Raman scattering in Fig. 3(b) can be explained with multiple phonon processes. We can interpret the structures at ~ 1464, 1638, 1696, and 1769 cm$^{-1}$ as multiphonon scatterings or multiple-order modes. The sum frequency of two modes, #6 (642 cm$^{-1}$) and #8 (826 cm$^{-1}$), is close to the mode at ~ 1464 cm$^{-1}$. Also, the mode at ~ 1638 cm$^{-1}$ is readily identified as the second-order mode of peak #8 (~ 826 cm$^{-1}$). The widths of the modes in Fig. 3(b) are very broad compared to those in Fig. 3(a). Similar behavior is also found in the multiple-order Raman scattering model of Martin [28]. According to the model, (i) multiple-order scattering involves an anharmonic potential with a three-phonon process (one localized mode and two band modes) resulting from the cubic anharmonicity, (ii) the scattering rate is evaluated with the second-order perturbation, and (iii) the result is proportional to the scattering order $N$ [28]. For example, the linewidth of the second-order scattering is expected to be a double that of the first-order scattering.

The most striking feature of Fig. 3(a) is that mode #6 becomes broad as Co substitutes at the Ni site. Note that mode #6 ($\sim 642$ cm$^{-1}$) and its multiphonon mode (at $\sim 1464$ cm$^{-1}$) are significantly broadened as a function of Co composition, $x$: especially, the mode at $\sim 1464$ cm$^{-1}$ is considerably overdamped in the cobalt end. It is intriguing to analyze this result in connection with the observed local symmetry change at the NVO end, i.e., the concurrent existences of Raman mode #6 and IR mode $b_9$ near the INS $\beta$-peak. The broadening and the overdamping of the modes are interpreted as a disappearance of Raman mode #6, for its lifetime is decreasing. Also, it suggests that the symmetry rule is restored as a function of $x$. Through our Raman results for NCVO, we suggest that the multiferroicity in NVO is related to the broken symmetry.

Then, why are our room-temperature vibrational spectroscopic results meaningful in the context of the multiferroicity of NVO occurring at low temperature? Previous studies suggest that the origin of the spontaneous polarization along the $b$-axis can be ascribed to one of the IR $B_{2u}$ modes, either the $b_9$ or the $b_4$ mode [18, 19]. Yildirim *et al.* chose the $b_9$ mode because it had the largest oscillator strength among $B_{2u}$ phonons and it could induce a significant dipole moment [18]. However, their room-temperature results did not show the soft mode behavior of the $b_9$ mode. Instead, Vergara *et al.* performed low-temperature measurements and found a softening behavior of $b_4$ mode. They also found a negligible change in the $b_9$ mode frequency during the HTI-LTI transition (ferroelectric-ordering transition) [19]. Our Raman measurements on NCVO provide similar conclusions through different approaches: (*i*) The $c$ mode (similar to the IR $b_4$ mode) vanishes at $x > 2.0$. (*ii*) The doublet modes, #6 (similar to the IR $b_9$ mode) and #7, become overdamped with increasing $x$. Our sequential measurements provide another trajectory for vanishing multiferroism and suggest that both IR modes ($b_4$ and $b_9$) are closely related to the ferroelectricity of NVO.

Additionally, our Raman results sensitively depend on the oxygen content in Co-rich samples ($x > 1.5$). Figure 4 depicts the oxygen sensitivity very clearly: the $\sim 853$ cm$^{-1}$ peak was detected only in oxygen-annealed cases, but not in argon-annealed ones. An oxygen deficiency or excess often leads to the appearance of forbidden Raman modes in oxides, e.g., SrTiO$_{3-\delta}$ [29] or CeO$_{2-\delta}$ [30].

# IV. CONCLUSIONS

The Raman modes of NCVO solid solutions were investigated. Co-substitution induces a volume expansion in the NCVO lattices such that phonon modes are softened through the mode Grüneisen parameter representing the doping-induced chemical pressure effect. The Raman spectrum of NCVO is critically compared with the INS spectrum and the IR phonon modes. As important modes for multiferroicity, the Raman modes (mode $c$, #6, and #7) are comparable to the INS peaks ($\alpha$ and $\beta$) and to the IR $B_{2u}$ modes ($b_4$ and $b_9$). We found that these modes disappeared with increasing Co-concentration and that these observations were closely correlated to the symmetry restoration and to the missing multiferroicity in CVO.

# ACKNOWLEDGMENTS

This work was supported by the National Research Foundation of Korea (NRF) grant funded by the Ministry of Education, Science, and Technology (MEST) (No. 2012-006641). IKJ was supported by the NRF grant funded by the MEST (No. 2012-0000345).

# REFERENCES

[1] W. Eerenstein, N. D. Mathur, and J. F. Scott, Nature **442**, 759 (2006).

[2] S.-W. Cheong and M. Mostovoy, Nat. Mater. **6**, 13 (2007).

[3] T. Kimura, T. Goto, H. Shintani, K. Ishizaka, T. Arima, and Y. Tokura, Nature **426**, 55 (2003).

[4] N. Hur, S. Park, P. A. Sharma, J. S. Ahn, S. Guha, and S.-W. Cheong, Nature **429**, 392 (2004).

[5] G. A. Smolenskii and I. E. Chupis, Sov. Phys. -Usp. **25**, 475 (1982).

[6] T. Kimura, J. C. Lashley, and A. P. Ramirez, Phys. Rev. B **73**, 220401 (2006).

[7] Y. Yamasaki, S. Miyasaka, Y. Kaneko, J.-P. He, T. Arima, and Y. Tokura, Phys. Rev. Lett. **96**, 207204 (2006).

[8] G. Lawes, A. B. Harris, T. Kimura, N. Rogado, R. J. Cava, A. Aharony, O. Entin-Wohlman, T. Yildirim, M. Kenzelmann, C. Broholm, and A. P. Ramirez, Phys. Rev. Lett. **95**, 087205 (2005).

[9] T. Kimura, G. Lawes, and A. P. Ramirez, Phys. Rev. Lett. **94**, 137201 (2005).

[10] N. Rogado, G. Lawes, D. A. Huse, A. P. Ramirez, and R. J. Cava, Solid State Commun. **124**, 229 (2002).

[11] G. Lawes, M. Kenzelmann, N. Rogado, K. H. Kim, G. A. Jorge, R. J. Cava, A. Aharony, O. Entin-Wohlman, A. B. Harris, T. Yildirim, Q. Z. Huang, S. Park, C. Broholm, and A. P. Ramirez, Phys. Rev. Lett. **93**, 247201 (2004).

[12] M. Kenzelmann, A. B. Harris, A. Aharony, O. Entin-Wohlman, T. Yildirim, Q. Huang, S. Park, G. Lawes, C. Broholm, N. Rogado, R. J. Cava, K. H. Kim, G. Jorge, and A. P. Ramirez, Phys. Rev. B **74**, 014429 (2006).

[13] R. Szymczak, M. Baran, R. Diduszko, J. Fink-Finowicki, M. Gutowska, A. Szewczyk, and H. Szymczak, Phys. Rev. B **73**, 094425 (2006).

[14] Y. Chen, J. W. Lynn, Q. Huang, F. M. Woodward, T. Yildirim, G. Lawes, A. P. Ramirez, N. Rogado, R. J. Cava, A. Aharony, O. Entin-Wohlman, and A. B. Hariss, Phys. Rev. B **74**, 014430 (2006).

[15] N. Qureshi, H. Fuess, H. Ehrenberg, T. C. Hansen, C. Ritter, K. Prokes, A. Podlesnyak, and D. Schwabe, Phys. Rev. B **74**, 212407 (2006).

[16] N. Qureshi, H. Fuess, H. Ehrenberg, B. Ouladdiaf, J. Rodríguez-Carvajal, T. C. Hansen,

Th. Wolf, C. Meingast, Q. Zhang, W. Knafo, and H. v. Löhneysen, J. Phys.: Condens. Matter **20**, 235228 (2008).

[17] A. B. Harris, T. Yildirim, A. Aharony, and O. Entin-Wohlman, Phys. Rev. B **73**, 184433 (2006).

[18] T. Yildirim, L. I. Vergara, J. Iñiguez, J. L. Musfeldt, A. B. Harris, N. Rogado, R. J. Cava, F. Yen, R. P. Chaudhury, and B. Lorenz, J. Phys.: Condens. Matter **20**, 434214 (2008).

[19] L. I. Vergara, J. Cao, N. Rogado, Y. Q. Wang, R. P. Chaudhury, R. J. Cava, B. Lorenz, and J. L. Musfeldt, Phys. Rev. B **80**, 052303 (2009).

[20] C. Sudakar, P. Kharel, R. Naik, and G. Lawes, Phylos. Mag. Lett. **87**, 223 (2006).

[21] F. Izumi and T. Ikeda, Mater. Sci. Forum **321-324**, 198 (2000).

[22] For comparison, in Ref. 10, $a = 5.931(3)$, $b = 11.374(3)$, and $c = 8.235(3)$ Å for NVO, and $a = 6.045(4)$, $b = 11.517(3)$, and $c = 8.316(3)$ Å for CVO.

[23] R. D. Shannon, Acta. Cryst. A **32**, 751 (1976).

[24] L. Vegard, Z. Phys. **5**, 17, (1921); Z. Kristallogr. **67**, 239 (1928).

[25] N. W. Ashcroft and N. D. Mermin, *Solid State Physics* (Saunders, Philadelphia, 1976), p. 493.

[26] F. D. Hardcastle, I. E. Wachs, H. Eckert, and D. A. Jeerson, J. Solid State Chem. **90**, 194 (1991).

[27] N. Hollmann, Z. Hu, T. Willers, L. Bohatý, P. Becker, A. Tanaka, H. H. Hsieh, H.-J. Lin, C. T. Chen, and L. H. Tjeng, Phys. Rev. B **82**, 184429 (2010).

[28] T. P. Martin, Phys. Rev. B **13**, 3617 (1976).

[29] D. A. Tenne, I. E. Gonenli, A. Soukiassian, D. G. Schlom, S. M. Nakhmanson, K. M. Rabe, and X. X. Xi, Phys. Rev. B **76**, 024303 (2007).

[30] J. R. McBride, K. C. Hass, B. D. Poindexter, and W. H. Weber, J. Appl. Phys. **76**, 2435 (1994).

**Figure Captions**

Fig. 1. IR transmission spectra of NVO and CVO using KBr pellets with concentrations of (a) 0.1 wt.%, (b) 0.5 wt.%, and (c) 1.0 wt.%.

Fig. 2. (a) Raman spectrum and (b) inelastic neutron scattering spectrum of NVO from Ref. 17. The Raman spectrum below 700 cm$^{-1}$ is multiplied by a scaling factor, as indicated. For a comparison, infrared modes from Ref. 18 are shown as lines in (b).

Fig. 3. Raman scattering spectra of $Ni_{3-x}Co_xV_2O_8$: (a) 0-1000 cm$^{-1}$ and (b) 1000-2000 cm$^{-1}$. The spectra were measured with laser power of 18 mW, except NVO. For NVO, 8 mW was used. The spectra were normalized to the strength of peak #8. (c) Center frequencies of selected modes *vs*. *x*. (d) The deviation of the center frequency *vs*. the deviation of volume with respect to NVO ($x = 0$): #2(○), #3(Δ), #6(□), and #8(◊).

Fig. 4. Raman spectra of $Ni_{3-x}Co_xV_2O_8$ ($x = 2.5$ and 3.0) samples annealed under two different gas environment (oxygen and argon).

**Table 1** Experimental values of the Raman mode frequencies for $Ni_{3-x}Co_xV_2O_8$ compounds. Values are in cm$^{-1}$. The persistent peaks are indexed with numbers, and the vanishing modes are written in alphabets. $\gamma_j$ is the mode Grüneisen parameter of the $j$-th mode.

| Peak | $x=0$ | 0.5 | 1.0 | 1.5 | 2.0 | 2.5 | 3.0 | $\gamma_j$ |
|---|---|---|---|---|---|---|---|---|
| #1 | 163 | 157 | 152 | 149 | 144 | 139 | 136 | 5.86 |
| a | 182 | 178 | 176 | 174 | 173 | | | |
| #2 | 211 | 205 | 202 | 196 | 194 | 187 | 179 | 4.65 |
| b | 256 | | | | | | | |
| c | 270 | 267 | 265 | 261 | 260 | | | |
| d | 323 | 319 | | | | | | |
| #3 | 347 | 342 | 340 | 333 | 331 | 327 | 320 | 2.29 |
| #4 | 402 | 397 | 394 | 391 | 389 | 386 | 384 | 1.47 |
| #5 | 463 | 453 | 456 | 451 | 452 | 450 | 450 | 1.06 |
| #6 | 642 | 636 | 628 | 627 | 624 | 621 | 619 | 1.23 |
| #7 | 675 | 672 | 670 | 668 | 667 | 667 | 666 | 0.46 |
| e | 803 | 797 | 803 | | | | | |
| #8 | 826 | 823 | 822 | 819 | 817 | 814 | 811 | 0.51 |

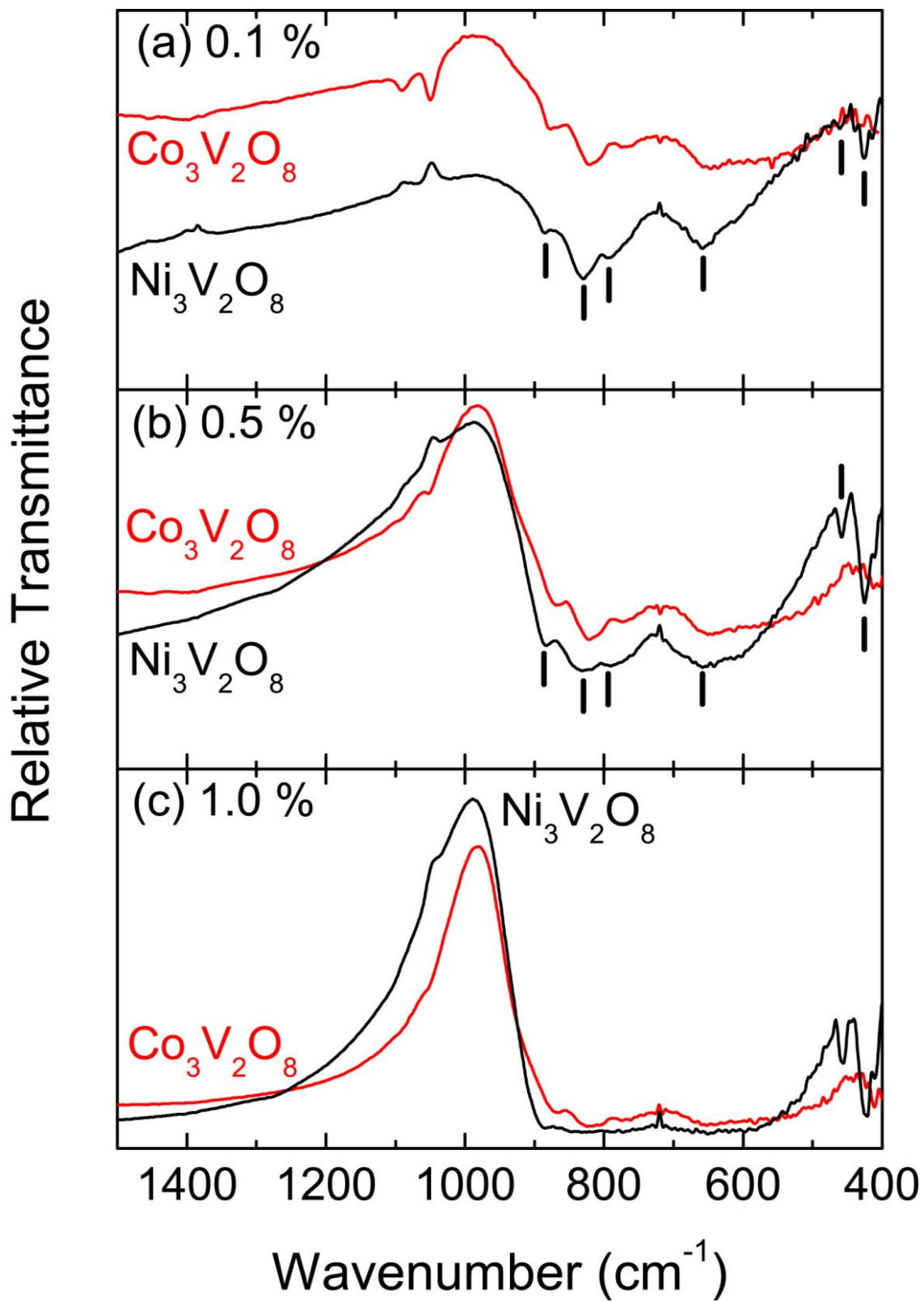

Fig. 1. (Color online) Yu-Seong Seo *et al*.

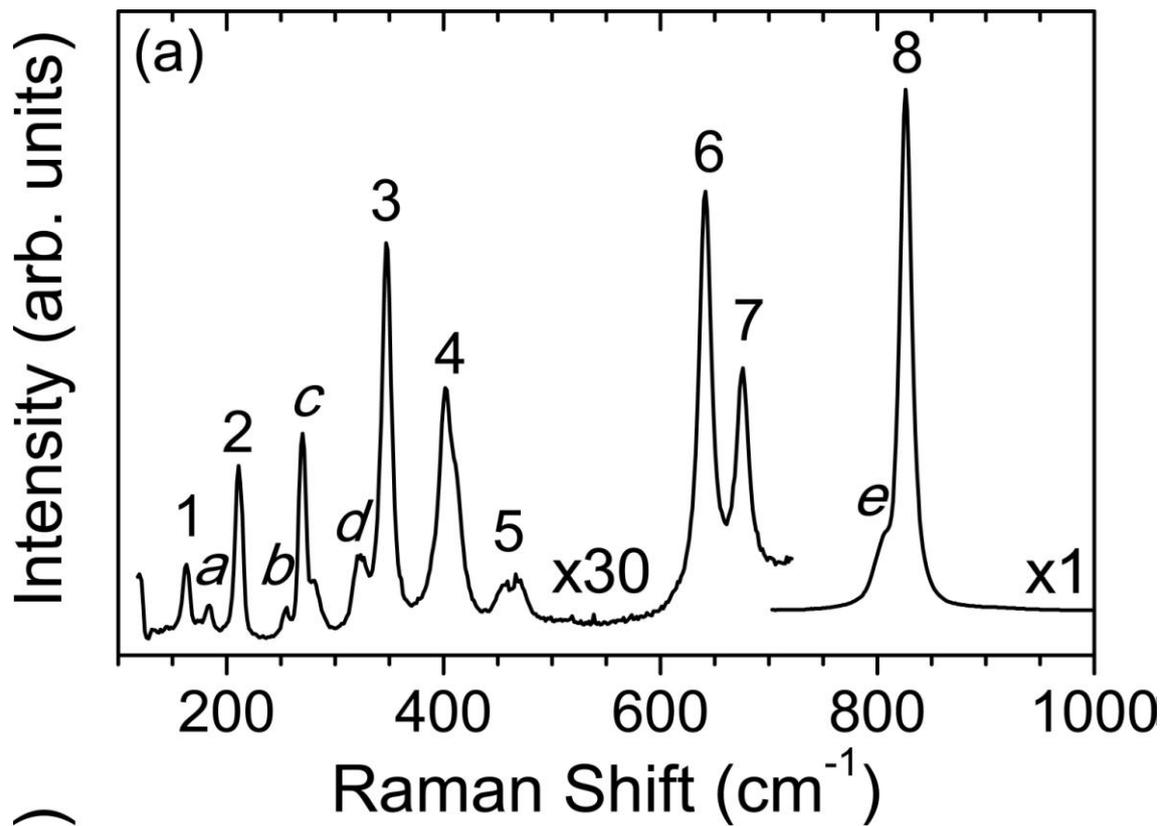

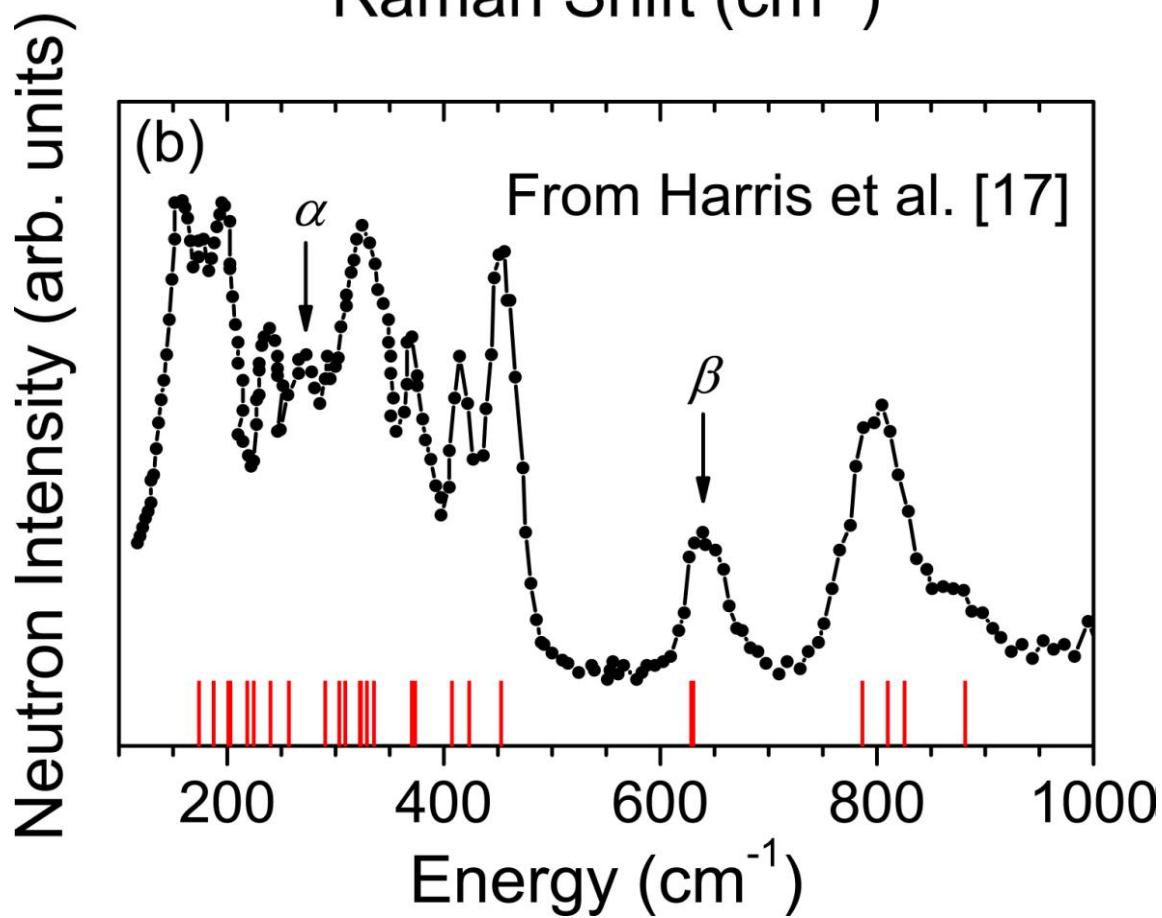

Fig. 2. (Color online) Yu-Seong Seo *et al*.

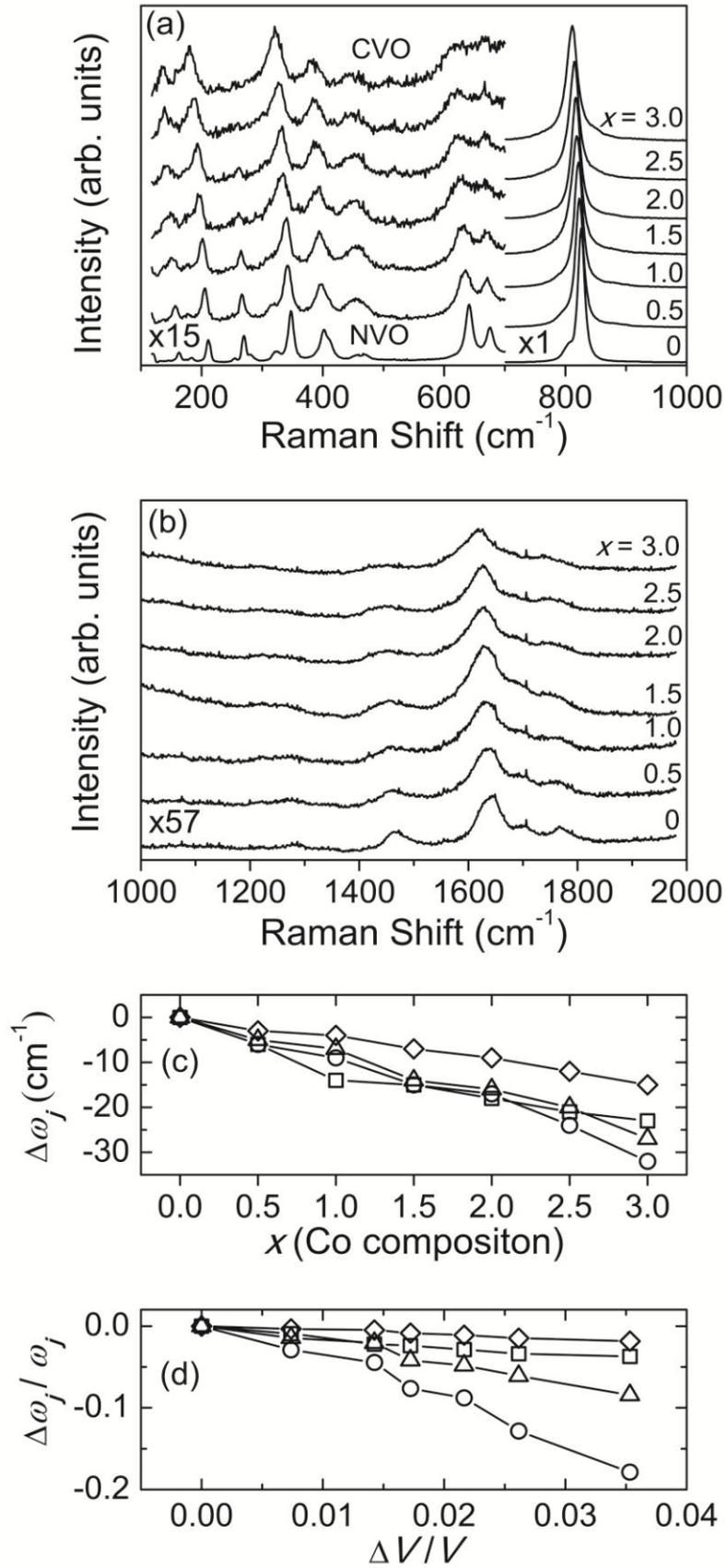

Fig. 3. Yu-Seong Seo *et al.*

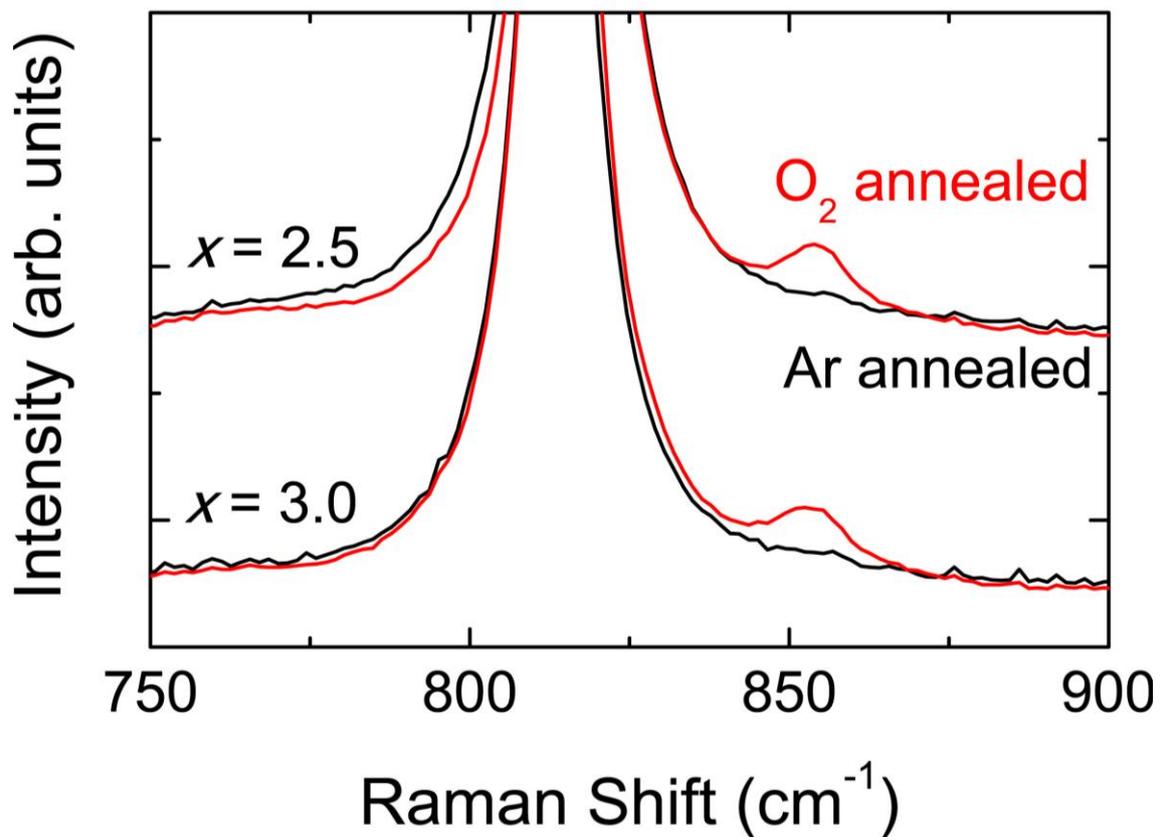

Fig. 4. (Color online) Yu-Seong Seo *et al*.